%% file: SPL-32939-2022_Report.tex
\documentclass[journal]{IEEEtran}

\onecolumn

\usepackage{lipsum}
\usepackage{multirow}
\usepackage{nicefrac}

\usepackage[boxed,vlined,ruled]{algorithm2e}
\SetAlCapNameFnt{\small}
\SetAlCapFnt{\small}
\usepackage[utf8]{inputenc} 
\usepackage[T1]{fontenc}
\usepackage{url}
\usepackage{ifthen}
\usepackage{fancybox}
\usepackage[usenames,dvipsnames,table]{xcolor}

\usepackage[scaled=.80]{beramono}

\usepackage{array,graphicx}
\usepackage{amsfonts,cite,euscript}
\usepackage[cmex10]{amsmath}
\usepackage[amsmath]{empheq}
\usepackage{cases,balance,cite, lineno}
\usepackage{booktabs,mathrsfs,bbm,amssymb,soul,lipsum}
\usepackage{amsthm}
\usepackage{dsfont}

\usepackage[inline, shortlabels]{enumitem}
\setlist{itemjoin ={,\enspace}, itemjoin* = {, and\enspace}}

\graphicspath{{./Figures/}}

\usepackage[percent]{overpic}
\newcommand{\bpara}[1]		{\medskip \noindent {\bf #1}}
\newcommand{\bparab}[1] 	{\noindent {\bf #1}}

\definecolor{modulo}{RGB}{72,0,255}
\definecolor{tof}{RGB}{255,0,63}

\usepackage{hyperref}
\hypersetup{
    colorlinks=true,
    linkcolor=black,
    filecolor=black,      
    urlcolor=black,
    citecolor = blue,
}

\theoremstyle{plain}
\newtheorem*{theorem*}{Theorem}
\newtheorem{theorem}{Theorem}

\newtheorem*{eg}{Example}

\usepackage{physics}
\usepackage{tikz}
\usepackage{mathdots}
\usepackage{yhmath}
\usepackage{cancel}
\usepackage{siunitx}
\usepackage{multirow}
\usepackage{gensymb}
\usepackage{tabularx}
\usepackage{extarrows}
\usepackage{booktabs}
\usetikzlibrary{fadings}
\usetikzlibrary{patterns}
\usetikzlibrary{shadows.blur}
\usetikzlibrary{shapes}

\usepackage{caption}

\def\Z					{\mathbb Z}
\def\R					{\mathbb R}

\def\iZ					{\in \mathbb Z}

\def\ind					{\mathbbmtt{1}}
\def\DE					{\stackrel{\rm{def}}{=}}

\def\usadc					{{\fontsize{11pt}{11pt}\selectfont\texttt{US}-\texttt{ADC}}}
\def\ussr					{{\fontsize{11pt}{11pt}\selectfont\texttt{US}-\texttt{SR}}}

\def\l						{\left(}
\def\r						{\right)}

\def\ind					{\mathds{1}}

\def\eg					{\emph{e.g.~}}
\def\ie					{\emph{i.e.}}

\def\tbmat					{\boldsymbol{\EuScript{T}}_{{\mat{b}}}^M  }
\def\tbmatL				{\boldsymbol{\EuScript{T}}_{{\mat{b}}}^{\ML}  }
\newcommand\TM[1]			{{\boldsymbol{\EuScript{T}}}_{{#1}}}

\def\dvmat					{\bs{\EuScript{D}}_{\hat{\bs{\varphi}}}}
\def\ddvmat				{\bs{\EuScript{D}}_{\hat{\bs{\bar{\varphi}}}}}
\def\ddvmatP				{\boldsymbol{\EuScript{D}}^{+}_{\hat{\bar{\bs{\varphi}}}}}
\def\rdc					{c_{\texttt{DC}}}

\def\qp					{\widehat{s}_p}
\def\qmat					{\widehat{\mathbf{s}}}
\def\qsmat					{\widetilde{\mathbf{s}}}

\newcommand\cin[2]			{\in \mathbb{C}^{#1 \times #2}}
\newcommand\rin[2]			{\in \mathbb{R}^{#1 \times #2}}
\newcommand\prm[1]		{\texttt{PRONY}\rob{#1}}

\newcommand\prest[3]		{\rob{\{{#1}_{#3}, {#2}_{#3}\}_{#3 = 0}^{\uppercase{#3}-1},\mat{h}}}
\newcommand\prestL[3]		{\rob{\{{#1}_{#3}, {#2}_{#3}\}_{#3 = 0}^{\color{black}\uppercase{#3}_\lambda-1},\mat{h}}}

\newcommand\bs[1]			{\boldsymbol{#1}}

\newcommand\eset[1]		{{\color{black}\mathcal{E}}_{P,#1}}
\newcommand\mcal[1]		{\mathcal{#1}}

\newcommand\rob[1]			{\l #1 \r}
\newcommand\fig[1]			{Fig.~\ref{#1}}

\newcommand{\sqb}[1]		{\left[ #1 \right]}

\newcommand{\ft}[1]			{\left[\kern-0.15em\left[#1\right]\kern-0.15em\right]}
\newcommand{\fe}[1]		{\left[\kern-0.30em\left[#1\right]\kern-0.30em\right]}
\newcommand{\flr}[1]		{\left\lfloor #1 \right\rfloor}

\newcommand{\MO}[1]		{\mathscr{M}_\lambda ({#1} )}

\newcommand{\RO}[1]		{\mathscr{R}_{#1}}

\newcommand{\mdft}[1]		{\hat{\bar{\mathbf{#1}}}}

\newcommand{\BL}[1]		{#1 \in \mathcal{B}_{\Omega}}

\newcommand{\EQc}[1]		{\stackrel{(\ref{#1})}{=}}

\newcommand{\mat}[1]		{\mathbf{#1}}

\newcommand{\ep}[1]		{\times 10^{#1}}

\def\tm					{{\color{black}\tau_m}}

\def\rto					{{\color{black}\leftarrow}}
\def\ML					{{\color{black}M_\lambda}}

\renewcommand\bar\underline
\renewcommand\hat\widehat
\renewcommand\geq\geqslant
\renewcommand\leq\leqslant
\renewcommand\Psi\ddvmat

\renewcommand\tilde\widetilde

\makeatletter
\def\moverlay{\mathpalette\mov@rlay}
\def\mov@rlay#1#2{\leavevmode\vtop{%
   \baselineskip\z@skip \lineskiplimit-\maxdimen
   \ialign{\hfil$\m@th#1##$\hfil\cr#2\crcr}}}
\newcommand{\charfusion}[3][\mathord]{
    #1{\ifx#1\mathop\vphantom{#2}\fi
        \mathpalette\mov@rlay{#2\cr#3}
      }
    \ifx#1\mathop\expandafter\displaylimits\fi}
\makeatother

\newcommand{\cupdot}{\charfusion[\mathbin]{\cup}{\cdot}}

\usepackage[most]{tcolorbox}
\newtcbtheorem{stp}{Step}%
{ 
    grow to left by=-0.2cm,grow to right by=0.45cm,
    bottom = -0.25cm,
    enhanced,frame empty,interior empty,
    colframe=black!60,
    borderline west={1pt}{1pt}{black},
    left=0.1cm,
    attach boxed title to top left={yshift=-2mm,xshift=-2mm},
    coltitle=black,
    fonttitle={\bf \small},
    colbacktitle=Goldenrod!15,
    boxed title style={boxrule=.5pt,sharp corners,size=small}}{Steps}

\ifCLASSINFOpdf
\else
\fi
\begin{document}

\title{Back in the US-SR: Unlimited Sampling and Sparse Super-Resolution with its Hardware Validation}

\author{Ayush~Bhandari

\thanks{A.~Bhandari's work is supported by the UK Research and Innovation council's \emph{Future Leaders Fellowship} program ``Sensing Beyond Barriers'' (MRC Fellowship award no.~MR/S034897/1).
Project page for (future) release of hardware design, code and data: \href{https://bit.ly/USF-Link}{\texttt{https://bit.ly/USF-Link}}.}
\thanks{A.~Bhandari is with the Dept. of Electrical and Electronic Engineering, Imperial College London, South Kensington, London SW7 2AZ, UK. (Email: \texttt{a.bhandari@imperial.ac.uk} or \texttt{ayush@alum.mit.edu}).}
%
\thanks{Manuscript submitted on February 23, 2022; accepted March 17, 2022. Date of publication March 23, 2022; date of current version May 3, 2022. }}

\markboth{\sf{IEEE Sig. Proc. Letters, Mar. 2022.}}%
{AB \MakeLowercase{\textit{et al.}}: Unlimited Sampling and Super-resolution}

\maketitle

\begin{abstract}  
The {Unlimited Sensing Framework} (USF) is a digital acquisition protocol that allows for sampling and reconstruction of high dynamic range signals. By acquiring modulo samples, the USF circumvents the clipping or saturation problem that is a fundamental bottleneck in conventional analog-to-digital converters (ADCs). In the context of the USF, several works have focused on bandlimited function classes and recently, a hardware validation of the modulo sampling approach has been presented. In a different direction, in this paper we focus on non-bandlimited function classes and consider the well-known  super-resolution problem; we study the recovery of sparse signals (Dirac impulses) from low-pass filtered, modulo samples. Taking an end-to-end approach to USF based super-resolution, we present a novel recovery algorithm (US-SR) that leverages a doubly sparse structure of the modulo samples. We derive a sampling criterion for the US-SR method. A hardware experiment with the modulo ADC demonstrates the empirical robustness  of our method in a realistic, noisy setting, thus validating its practical utility.
\end{abstract}
\begin{IEEEkeywords}
Analog-to-digital, modulo sampling, Shannon sampling, spectral estimation, Prony's method, super-resolution.
\end{IEEEkeywords}

\IEEEpeerreviewmaketitle
\bigskip
\bigskip

{\centering

{\color{red} This paper was published in IEEE Signal Processing Letters, vol.~29, March 2022. \\ 
\href{https://doi.org/10.1109/LSP.2022.3161865}{\color{blue}\texttt{DOI: 10.1109/LSP.2022.3161865}}.}

}

\bigskip
\bigskip
\tableofcontents

\newpage
\linespread{1.2}

\section{Introduction}
\begin{figure}[!t]
\centering
\begin{overpic}[width=0.65\columnwidth]{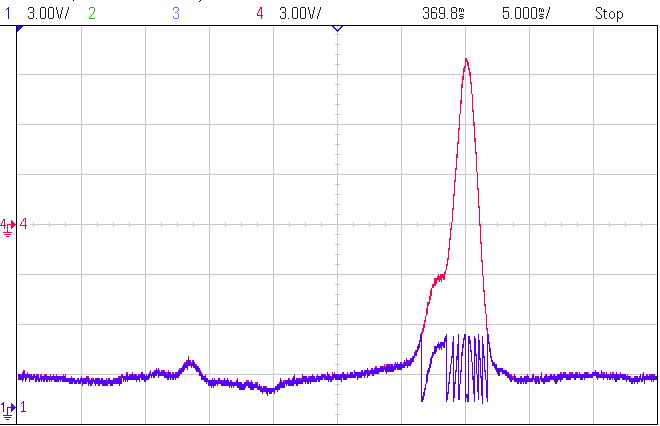}
\put (4,50) {\colorbox{white}{\sf{\bfseries\fontsize{8pt}{8pt}\selectfont {\color{tof}------ $\pmb{g\rob{t}}$ }}}}
\put (10,46) {\colorbox{white}{\sf{\bfseries\fontsize{7pt}{7pt}\selectfont {\color{tof} Low-Pass Filtered Spikes ($K = 2$)}}}}
\put (10,42) {\colorbox{white}{\sf{\bfseries\fontsize{7pt}{7pt}\selectfont {\color{tof} (Time-Resolved Measurements \cite{Bhandari:2016:J})}}}}
\put (4,16) {\colorbox{white}{\sf{\bfseries\fontsize{7pt}{7pt}\selectfont {\color{modulo}------ $\pmb{\MO{g\rob{t}}}$}}}}
\put (10,12) {\colorbox{white}{\sf{\bfseries\fontsize{7pt}{7pt}\selectfont {\color{modulo}Noisy Modulo Signal}}}}
\end{overpic}
\caption{Oscilloscope screenshot of filtered spikes measured using a time-resolved imaging sensor \cite{Bhandari:2016:J} and the corresponding modulo signal obtained via our prototype modulo ADC \cite{Bhandari:2021:J}. For recovery, see Section~\ref{sec:HEV} and \fig{fig:exp}.}
\label{fig:demo}
\end{figure}
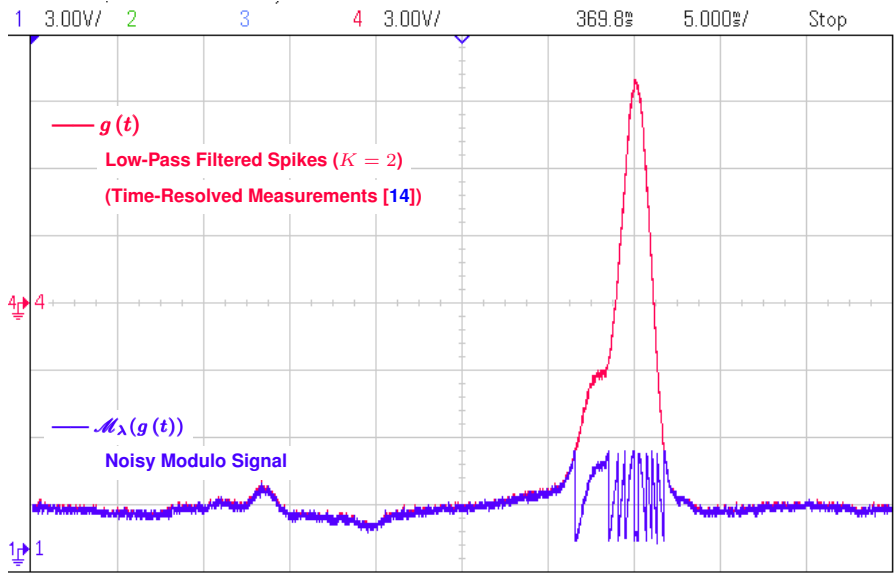

\IEEEPARstart{T}{he} {Unlimited Sensing Framework} (USF) \cite{Bhandari:2017:C,Bhandari:2018:Ca,Bhandari:2018:C,Bhandari:2019:C,Bhandari:2020:Pata,Bhandari:2020:Ja,Bhandari:2021:J,FernandezMenduina:2021:J,Florescu:2022:J} has been recently proposed in the literature to circumvent signal clipping or saturation problem in digital acquisition. 
Conventional analog-to-digital converters (ADCs) have a fixed dynamic range (DR). This poses a fundamental limitation; in real-world scenarios when an input signal, say $g\rob{t}, t\in\R$, exceeds the ADC's DR $\rob{\lambda}$, the resulting samples are clipped or saturated \cite{Abel:1991:C,Esqueda:2016:J,Olofsson:2005}. 
The USF goes beyond the sequential \emph{capture first, process later} pipeline by capitalizing on a joint design of hardware and algorithms, which is the very essence of \emph{computational sensing} \cite{Bhandari:2022:Book}. 
A joint design is achieved by,

\begin{enumerate}[leftmargin = 20pt,label = $\bullet$,itemsep = 5pt]
\item {\bf Hardware Based Modulo Encoding.} Modulo folding before sampling ensures that the high dynamic range (HDR) input does not exceed the ADC's DR. This is accomplished by injecting a centered modulo non-linearity defined by, 
\begin{equation}
\label{map}
\mathscr{M}_{\lambda}:g \mapsto 2\lambda \left( {\fe{ {\frac{g}{{2\lambda }} + \frac{1}{2} } } - \frac{1}{2} } \right), \ \  \lambda >0
\end{equation}
where $\ft{g} \DE g - \flr{g}$ and $\flr{g}  = \sup \left\{ {\left. {m \in \mathbb{Z}} \right|m \leqslant g} \right\}$ (floor function). 
To bridge the gap between theory and practice, we have developed a modulo ADC \cite{Bhandari:2021:J}, \emph{viz}. the \usadc, providing the first hardware validation of the USF approach. In particular, we have shown that signals as large as $24\lambda$ can be recovered in practice. 
The output of the \usadc~is shown in \fig{fig:demo}. Clearly, $|\MO{g}|\leq \lambda$.

\item {\bf Algorithm Based Decoding.} Given folded samples $y\sqb{n} = \MO{g\rob{nT}}, T>0$, algorithmic decoding is performed to solve the inverse problem of recovering $\gamma\sqb{n} = g\rob{nT}$ using mathematically guaranteed recovery algorithms. 
\end{enumerate}

{For a finite-energy function with maximum frequency $\Omega$ (rad/s) denoted by $\BL{g}$}, we have proved a \emph{Shannon-Nyquist} like  principle; a constant factor oversampling \ie~$ T\leq \nicefrac{1}{2\Omega e}$, $e \approx 2.718$ (Euler's constant), independent of $\rob{\lambda}$, suffices recovery \cite{Bhandari:2017:C,Bhandari:2019:C,Bhandari:2020:Ja}. Conventional wisdom is that bandlimited signals can not be recovered if the spectrum is aliased. Counter-intuitively, despite $\MO{g}\not\in\mathcal{B}_\Omega$ leading to aliased spectrum, bounded {time-bandwidth} product $\Omega T\leq \nicefrac{1}{2 e}$ guarantees {inversion} of $\MO{\cdot}$. A similar inversion method applies to non-bandlimited functions, \eg spline spaces \cite{Bhandari:2020:C}. With the knowledge of finitely many unfolded samples, recovery based on Nyquist rate sampling has been demonstrated in \cite{Romanov:2019:J}. For other USF related recovery approaches, we refer to \cite{Musa:2018:C,Rudresh:2018:C,Gan:2020:C,Shah:2021:J,Prasanna:2021:J}.

\bpara{Motivation.} 
Sparse signal recovery from low-pass filtered samples is an important  topic backed by decades of interdisciplinary progress. Related topics include, %
\begin{enumerate*}[label = \emph{\alph*})]
\item Tauberian approximation \cite{Figueiredo:1982:J}
\item Time-delay estimation \cite{Kirsteins:1987,Fuchs:1999:J,Gedalyahu:2010:J}
\item Super-resolution \cite{Donoho:1992:J,Candes:2013:J,Bhaskar:2013:J}
\item Sparse deconvolution \cite{Li:2000:J}
\item Finite-rate-of-innovation sampling \cite{Vetterli:2002:J,Blu:2008:J}.
\end{enumerate*}

Beyond the wide applicability of the SR model (cf.~various hardware experiments \cite{Bhandari:2014:Cb,Bhandari:2016:Cb,Bhandari:2017:Ca}), in certain applications, physical limits of the hardware impose a sampling time that is much larger than the time-scale of features to be resolved. This is particularly the case with time-resolved imaging\cite{Bhandari:2016:J,Bhandari:2014:Cb,Bhandari:2016:Cb,Bhandari:2020:J}, see \fig{fig:demo}. This is the setup considered in Section~\ref{sec:HEV}, where recovering \emph{echoes of light} at their time-scale is impractical and necessitates SR. At the same time, HDR scene reflectivity may lead to demanding DR considerations, as is the case with imaging \cite{Bhandari:2020:C}.

These above aspects clearly motivate the problem of super-resolution from modulo samples; For practical scenarios motivating this problem cf.~Fig.~1 in \cite{Bhandari:2018:Ca}. The initial approach in \cite{Bhandari:2018:Ca} adopts a sequential recovery method; (a) modulo samples are first unfolded yielding the usual samples, and then, (b) existing SR methods can be used. The recovery in \cite{Bhandari:2018:Ca} imposes practical limitations. Firstly, sparse priors are not exploited for unfolding, this results in highly demanding oversampling factors. Secondly, the reconstruction relies on inversion of higher order differences and this is highly sensitive to noise. Here, we present a novel method that does not suffer with such limitations and its utility as validated via the \usadc.

\bpara{Contributions.} Our main contribution is a novel SR method for USF, \emph{viz.} \ussr, and its hardware validation via
time-resolved imaging based experiment where the super-resolution model arises naturally. Key features of our work include, 
\begin{enumerate}[leftmargin = 30pt,label = $\blacktriangleright$]
  \item \ussr~is direct (hence efficient) in the sense that unfolding step in \cite{Bhandari:2018:Ca} is not needed. Also, \ussr~is \emph{agnostic} to $\lambda$.
  \item \ussr~is backed by theoretical guarantees that enable recovery at potentially lower sampling rates (than \cite{Bhandari:2018:Ca}).
  \item We use the \usadc \cite{Bhandari:2021:J} to validate our method on real experiments based on time-resolved imaging sensor data \cite{Bhandari:2016:J,Bhandari:2020:J}, in Section~\ref{sec:HEV}. This is motivated by two reasons, 
\begin{enumerate}[leftmargin =40pt,label = \roman*)]  
\item Time-resolved imaging is a significant research area. Our work gives a sense about realistic performance of \ussr~with a clear SR application \cite{Bhandari:2016:Cb} in mind.
\item The experiment also establishes the empirical robustness of \ussr~in the presence of system noise \eg quantization errors and additive Gaussian noise.   
\end{enumerate}  
\end{enumerate}

\section{Problem Setup}
We define our $K$-sparse signal to be super-resolved as, 
\begin{equation}
\label{sod}
{s_K}\left( t \right) \DE \sum\limits_{k = 0}^{K - 1} {{c_k}\delta \left( {t - {t_k}} \right)}, \qquad \sum\limits_k |c_k| < \infty.
\end{equation}
Conventional, pointwise samples arising from low-pass projections of the $K$-spikes with kernel $\varphi$ are given by, 
\begin{equation}
\label{gbl}
g\left( {nT} \right) = \left\langle {{s_K},\overline \varphi  \left( { \cdot  - nT} \right)} \right\rangle  = {\left. {\left( {{s_K}*\varphi } \right)\left( t \right)} \right|_{t = nT}}
\end{equation}
where $\BL{\varphi}$ is known, $\overline{\varphi}\rob{t} = \varphi\rob{-t}$ and $\left\langle {a,b} \right\rangle  = \int {a\left( x \right){b^*}\left( x \right)dx}$ denotes the $L^2$ inner-product. We will assume that the kernel $\varphi$ is $\tau$-periodic \cite{Vetterli:2002:J,Candes:2013:J} with $ \left\{ {{t_k}} \right\}_{k = 0}^{K - 1} \in \left[ {0,\tau } \right)$.
Our goal is to recover $s_K$ given the modulo samples, 
\begin{equation}
\label{yn}
y\sqb{n} = {\left. \MO{g\rob{t}} \right|_{t = nT}}, \qquad n = 0,\dots,N-1.
\end{equation}
In \cite{Bhandari:2018:Ca}, it is shown that,
\[ 
T\leq \frac{1}{2\Omega e} \mbox{ and } N\geq \rob{2K+1} + \frac{7\beta_g}{\lambda}, \quad \beta_g \geq ||g||_\infty
\] 
suffices recovery, which is based on the inversion of finite-differences of order $L = \left\lceil {\log \left( \nicefrac{\lambda} {\beta _g} \right)/\log \left( {T\Omega e} \right)} \right\rceil$. This creates a practical challenge---higher order differences are \emph{unstable} in the presence of perturbations. Even for nominal values of $L$, the reconstruction is highly ill-posed. 

\section{Recovery via Double Sparsity}

\bparab{Simplifying Measurements.} Since $\varphi\rob{t}$ is $\tau$-periodic, it admits a Fourier Series (FS) expansion, $\varphi \left( t \right) = \sum\nolimits_{p \iZ} {{{\widehat \varphi }_p}{e^{\jmath p {\omega _0}t}}} $, $\omega_0 = \nicefrac{2\pi}{\tau}$ where {${\widehat \varphi _p} = {\tau ^{ - 1}}{\left\langle {\varphi \left( \cdot \right),{e^{\jmath p {\omega _0}\cdot}}} \right\rangle _{\left[ {0,\tau } \right]}}$} are the FS coefficients. With $\BL{\varphi}$, we have $\widehat{\varphi}_p = 0, |p|>P = \left\lceil {\nicefrac{\Omega }{{{\omega _0}}}} \right\rceil$ (bandlimitedness). We can now simplify $g\rob{t}$ using,
\begin{align}
\label{GFT}
  g\left( t \right) 
  &\EQc{gbl} \sum\limits_{k = 0}^{K - 1} {{c_k}\varphi \left( {t - {t_k}} \right)}  \hfill \\
  & = \sum\limits_{p \leqslant \left| P \right|} {\rob{{\widehat \varphi }_p{\qp}} {e^{\jmath p{\omega _0}t}}},   
\qquad {\qp} = \sum\limits_{k = 0}^{K - 1} {{c_k}{e^{ - \jmath p{\omega _0}{t_k}}}}. \notag
\end{align}
In vector-matrix notation, the samples $\gamma\sqb{n}=g\rob{nT}$ read, 
\begin{equation}
\bs{\gamma} \EQc{gbl} \mat{U}^* \dvmat \qmat,
\label{gvec}
\end{equation} 
\begin{enumerate}[label = ---$\bullet$, leftmargin = 30pt]
  \item $\bs{\gamma}  \rin{N}{1}$ is the vector of samples with $\sqb{\bs{\gamma}}_n = g\rob{nT}$.
  \item $\mat{U}  \cin{N}{(2P+1)}$ is the Discrete Fourier Transform (DFT) matrix with element $\sqb{\mat{U}}_{n,p} = {{e^{-\jmath p{\omega _0}nT}}}$ with complex-conjugate denoted by $\sqb{\mat{U}}_{n,p}^*$.
  \item $\dvmat \cin{(2P+1)}{(2P+1)}$ is a diagonal matrix with FS coefficients of the kernel on the diagonal, 
\[
\sqb{\dvmat}_{n,n} =
\begin{cases}
\widehat\varphi_n & n \in \sqb{0,P} \cup \sqb{N-P,N-1} \\
0 & n \in \sqb{P+1,N-P-1} 
\end{cases}.
\]  
  
\item $\qmat\cin{(2P+1)}{1}$ is a vector of exponentials parameterized by the unknown $K$-sparse signal $s_K$, \ie~$\sqb{\qmat}_p = \widehat{s}_p$.
\end{enumerate}

\bpara{Towards the Doubly Sparse Structure of Modulo Samples.} {Modulo decomposition yields $g = \MO{g} + \RO{g}$ \cite{Bhandari:2017:C} where $t\in\sqb{0,\tau}, \RO{g}\rob{t} = \sum\nolimits_{m=0}^{\ML-1} {\mu_m{\ind_{{\mathcal{D}_m}}}\left( t \right)}$, $\mu_m\in 2\lambda\mathbb{Z}$ is the \emph{residue}, $\ind_\mcal{D}\rob{t} = 1, t\in\mcal{D}$ and $0$ elsewhere, with disjoint union $\cupdot_{m} {{\mathcal{D}_m}} = \sqb{0,\tau}$}, {and $\ML$ is the total number of folds induced by $\MO{\cdot}$}. The non-ideal case \cite{Bhandari:2021:J} when $\mu_m\in \R$ is also covered here. The decomposition applies to samples, $\bs{\gamma} = \mat{y} + \mat{r}$,  $\sqb{\mat{r}}_n =\RO{g}\rob{nT} $. For any $\mat{a}\rin{N}{1}$, let us define its first difference, $\bar a \sqb{n} \DE \rob{\Delta a} \sqb{n} \equiv a\sqb{n+1}-a\sqb{n} \Leftrightarrow \bar{\mat{a}} =  \bs{\Delta}\mat{a} $
where $\bs{\Delta} \rin{(N-1)}{N}$ is the difference matrix. Then, we have, $\bar{\mat{y}} = \bar {\bs{\gamma}} - \bar{\mat{r}}$ and the residue simplifies to,  
\begin{align}
\bar r \sqb{n} =	\sum\limits_{m=0}^{\ML-1} {\mu_m\delta\sqb{{nT - {\tm}}}},\ \  \tm \in \rob{T\Z}\cap \left[0,\tau\right)
\label{eq:spike}			
\end{align}
an unknown \emph{sparse signal} characterized by $2\ML$ unknowns, $\{ \mu_m,\tm \}_{m = 0}^{\ML-1}$ {and where $\delta\sqb{\cdot}$ is the Kronecker delta}. Finally, modulo samples are encoded as a \emph{doubly sparse} representation, 
\begin{equation}
\bar{\mat{y}} \equiv \bs{\Delta}\mat{y} = \bs{\Delta}\rob{{\bs{\gamma}} - {\mat{r}}} \EQc{gvec} \rob{\bs{\Delta}\mat{U}^* \dvmat} \qmat  - \bar{\mat{r}}
\label{yds}
\end{equation}
where $\bar{\mat{r}} =\bs{\Delta}\mat{r}$ is a $\ML$-sparse residual and $s_K$ is the $K$-sparse signal encoded as $\sqb{\qmat}_p = \sum\nolimits_{k = 0}^{K - 1} {{c_k}{e^{ - \jmath {\omega _0}p{t_k}}}}$. By mapping \eqref{yds} into Fourier Domain, we can encode $\bar{\mat{r}}$ as a trigonometric polynomial, this is similar in spirit to how $s_K$ is encoded via $\qmat$. Let us define the DFT of a vector $\bar{\mat{a}}\rin{(N-1)}{1}$ by $\mdft{a} = \mat{V}\bar{\mat{a}}$, $\sqb{\mat{V}}_{\ell,n} = e^{-\jmath \frac{2\pi}{N-1} \ell n}$, $0 \leq n \leq N-2$. Then we have, 
\begin{equation}
\mdft{y}  =  {\mat{V} {\bs{\Delta}\mat{U}^* \dvmat} \qmat}  - \mdft{r},  \ \  
\sqb{\mdft{r}}_\ell= \sum\limits_{m=0}^{\ML-1} {{\mu _m}{e^{ - \jmath \left( {\frac{{2\pi }}{{N - 1}}} \right)\frac{\ell }{T}{\tm}}}}. 
\label{yft}
\end{equation}
In the above, $\bs{\Psi} \DE \mat{V} {\bs{\Delta}\mat{U}^* \dvmat}$ is a \textit{spectrum shaping} filter arising from the first difference of the kernel $\varphi$ and $\hat{\bar{\bs{\gamma}}} = \bs{\Psi}\qmat$.

\bpara{Leveraging Double Sparsity for Super-resolution.} Given \eqref{yft}, our goal is to isolate $\qmat$ so that the unknowns $\left\{ {{c_k},{t_k}} \right\}_{k = 0}^{K - 1}$ can be estimated via spectral estimation methods \cite{Wolf:1983:J}. {Given $\mdft{y} = \mdft{\bs{\gamma}}-\mdft{r}$, the presence of $\BL{g}$ (cf.~\eqref{GFT},\eqref{gvec}) implies that} only $\rob{2P+1}$ out of $\rob{N-1}$ values of $\mdft{\bs{\gamma}}$ are non-zero while $\mdft{r}$ contributes to all of the $\rob{N-1}$ values. Hence, we will first isolate $\mdft{r}$. A similar insight was leveraged in our very recent work \cite{Bhandari:2021:J} but the key difference here is that our goal is to recover a non-bandlimited signal, $s_K$ in \eqref{sod} instead of $\BL{g}$. Below, we list the steps of our recovery method. \smallskip

\begin{figure}[!t]
\centering
\includegraphics[width=0.65\textwidth]{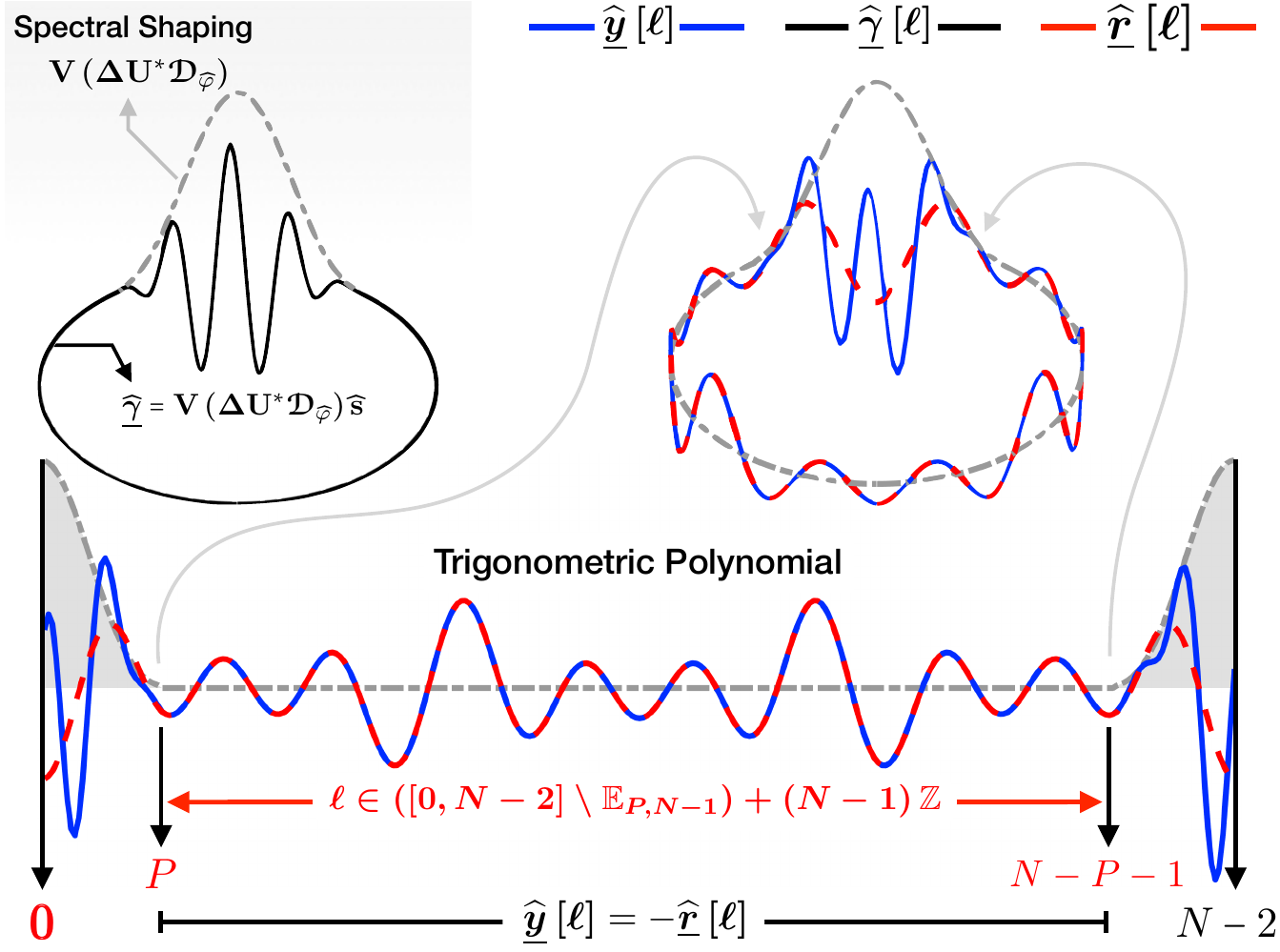}
\caption{Fourier domain partitioning of modulo samples as given by \eqref{eq:FP}.}
\label{fig:FT}
\end{figure}

\begin{stp}{Fourier Domain Partitioning.}{ex:stp0}
Let $\eset{N} = \left[ {0,P} \right] \cup \left[ {N - P,N - 1} \right]$ be the set on which $\widehat{\bs{\gamma}} = \mat{V}  {\mat{U}^* \dvmat} \qmat$ is supported with $\left| \eset{N} \right| = 2P + 1$, the bandwidth of $\varphi$ encoded in $\dvmat$.
Since $\BL{g}$, as shown in \fig{fig:FT}, we can partition the $\rob{N-1}$ Fourier coefficients of $\bs{\Delta}\mat{y}$ \ie~$\mdft{y}\cin{(N-1)}{1}$, on the circle as follows, 
\begin{equation}
\label{eq:FP}
\sqb{\mdft{y}}_\ell = 
\begin{cases}
\sqb{\bar{\bs{\gamma}} - \bar{\mat{r}}}_\ell  & 	\ell \in \eset{N-1}	\\
-\sqb{\mdft{r}}_\ell					 & 	\ell \in \left( {\left[ {0,N - 2} \right] \setminus \eset{N-1}} \right) 
\end{cases}.
\end{equation}
\end{stp}

\begin{stp}{Recovering Residue Samples via $\mdft{r}$.}{ex:stp1}
{Having isolated $\sqb{\mdft{r}}_\ell$, $\ell\in\mcal{L}_r = \rob{\left[ {0,N - 2} \right] \setminus \eset{N-1}} $, we obtain $\{ \mu_m,\tm \}_{m = 0}^{\ML-1}$ \eqref{yft} using Prony's method \cite{Figueiredo:1982:J}. Given $
{\left[ {\mathbf{b}} \right]_\ell } = \sum\nolimits_{m = 0}^{M - 1} {{\alpha_m}{e^{-\jmath {\nu_m}\ell }}}$, the symbolic representation $\prm{\mat{b}}$ provides both $\{ \alpha_m,\nu_m \}_{m = 0}^{M-1}$ and $\mat{h}\cin{(M+1)}{1}$}, the filter that \emph{annihilates} $\mat{b}\cin{L}{1}$ \cite{Blu:2008:J} \ie~$\rob{h*b}\sqb{\ell} = 0$ or $\mat{h} \in \ker ({\tbmat})$ via the matrix, 
\begin{equation*}
\tbmat \DE  \begin{bmatrix} 
  {{{\left[ {\mat{b}} \right]}_0}}&{{{\left[ {\mat{b}} \right]}_{ - 1}}}& \cdots &{{{\left[ {\mat{b}} \right]}_{ - M}}} \\ 
   \vdots & \vdots & \ddots & \vdots  \\ 
  {{{\left[ {\mat{b}} \right]}_{M - 1}}}&{{{\left[ {\mat{b}} \right]}_{M - 2}}}& \cdots &{{{\left[ {\mat{b}} \right]}_{ - 1}}} 
\end{bmatrix} \cin{M}{(M+1)}.
\label{eq:tmat}
\end{equation*}
{Note that $\mathsf{rank}(\tbmat) = M$ $\rob{\forall n,m, \nu_m\not = \nu_n}$ because when the same is constructed via $\sqb{\mat{u}_m}_\ell = e^{-\jmath {\nu_m}\ell}$, its rank is one.} We compute $\mat{h} \in \ker ({\tbmat}) \Longrightarrow \tbmat\mat{h} = \bs{0}$ whenever $L\geq 2M$. Similarly, with $\left| \mcal{L}_r \right| = \rob{N-2P-2}\geq 2\ML$, we obtain $\prestL{\mu}{\tau}{m} \rto \prm{\mdft{r}}$. 
\end{stp}

\begin{stp}{Recovering Fourier Samples of $\bs{\Delta\gamma}$ on $\eset{N-1}$.}{ex:stp2}
Having estimated $\mdft{r}$ from above, we obtain the Fourier samples, 
$\sqb{\mdft{\bs\gamma}}_\ell = \sqb{\mdft{y}}_\ell+ \sqb{\mdft{r}}_\ell$ at indices  
$\ell \in \eset{N-1}$.
\end{stp}

\begin{stp}{Fourier Deconvolution.}{ex:stp3}
With $2P+1$ values of $\hat{\bar{\bs{\gamma}}} = \mat{V}  {\bs{\Delta}\mat{U}^* \dvmat} \qmat$ known from above, it remains to estimate $s_K$ \eqref{sod}. First, we obtain, 
\begin{equation}
\label{sest}
\qsmat = \ddvmatP\mdft{\bs{\gamma}},
\ \ \quad \bs{\Psi} = \mat{V} {\bs{\Delta}\mat{U}^* \dvmat}
\end{equation}
where $\ddvmatP$ is the pseudo-inverse of $ \bs{\Psi} $. This operation reshapes or deconvolves $\sqb{\mdft{\bs\gamma}}_\ell$,  $\ell \in \eset{N-1}$ (cf.~\fig{fig:FT}) provided that $\widehat{\varphi}_p, |p|\leq P = \left\lceil {\nicefrac{\Omega }{{{\omega _0}}}} \right\rceil $ do not vanish. 
\end{stp}

\begin{stp}{Recovering the Unknown DC or $\bs{0}$ Frequency.}{ex:stp4}
Note that ${ \hat{\bar{\varphi}}_{\ell=0} }=0$ (due to $\Delta\varphi$ in \eqref{sest}) creating a blow-up in $\ddvmatP$ at $\ell=0$. Hence $\sqb{\qsmat}_0$ remains unknown. To estimate it, we assign $\sqb{\qsmat}_0$ an arbitrary value, \eg $\sqb{\qsmat}_0 = \sqb{\qsmat}_1$ and model $\widetilde{s}_{\ell} = \hat{s}_{\ell} + \rdc\delta\sqb{\ell}$ where $\hat{s}_{\ell}$ is defined in \eqref{GFT}. %
Since ${h*\hat{s}}=0$ (cf.~Step 2, above), we estimate $\rdc$ via the eigenvalue problem $\TM{\tilde{\mat{s}}} \mat{h} = \rdc \mat{h}$ where $\TM{\tilde{\mat{s}}} \cin{(K+1)}{(K+1)}$ (one extra row in $\tbmatL$) and the solution exists whenever $2K+1$ contiguous values of $\tilde{\mat{s}}$ are known, or $P\geq K$.
\end{stp}

\begin{stp}{Recovering Sparse Signal via $\mdft{s}$.}{ex:stp5}
Defining $\sqb{\hat{\mat{s}}}_0 = \rob{\sqb{\widetilde{\mat{s}}}_0 - \rdc}$, we use $\prm{\hat{\mat{s}}}$ to obtain $\prest{c}{t}{k}$ in \eqref{sod} where $\mat{h}$ is given in Step 5.
\end{stp}

\bpara{Recovery Condition.} Due to the parametric form of the underlying signals, there is an interplay of sampling time $\rob{T}$ and the  sample size $\rob{N}$. Step 2 requires $N\geq 2\rob{P+\ML + 1}$ while Step 5 enforces $P\geq K$. With $\tau = NT$, the recovery condition is,
$T\leq \nicefrac{\tau}{2\rob{P+\ML + 1}}$
with
$P = \left\lceil {\nicefrac{\Omega }{{{\omega _0}}}} \right\rceil \geq K$. Our result is formalized in the following theorem.

\begin{theorem} 
Suppose that we are given $N$ modulo samples $y\sqb{n} = \MO{g\rob{nT}}$, $T>0$ folded at most $\ML$ times, where $g = \rob{s_K*\varphi}$ and where ${s_K}\left( t \right) = \sum\nolimits_{k = 0}^{K - 1} {{c_k}\delta \left( {t - {t_k}} \right)}$ is an unknown $K$-sparse signal and $\BL{\varphi}$, is a known, $\tau$-periodic, kernel. Then, a sufficient condition for recovery of $s_K$ from $\{y\sqb{n}\}_{n=0}^{N-1}$ is that $T\leq\nicefrac{\tau}{N}$ and $N\geq 2\rob{K+\ML+1}$.
\end{theorem}

\setlength{\textfloatsep}{7pt}
\begin{algorithm}[!t]
\SetAlgoLined
{\bf Input:} $\{y\sqb{n},\MO{\varphi\rob{nT}}\}_{n=0}^{N-1}$, $\tau$, and $P$.\\
\KwResult{$K$-sparse signal $s_K\rob{t}$.}

\begin{enumerate}[label = $\arabic*)$,leftmargin=40pt,itemsep=1pt]
\item Compute $\mdft{y} = \mat{V} \mat{\bar{y}}$ (DFT) as in \eqref{yft} where $\bar{\mat{y}} = \bs{\Delta}\mat{y}$ \eqref{yds}.
\item Define $\sqb{\mat{z}}_\ell = -\sqb{\mdft{y}}_\ell$, $\ell \in \mcal{L}_r = \rob{\left[ {0,N - 2} \right] \setminus \eset{N-1}} $.
\item Estimate $\ML$ by forming a Toeplitz or Hankel matrix from $\sqb{\mat{z}}_\ell$ (cf.~\cite{Bhandari:2021:J}) and then thresholding using \emph{second order statistic} of eigenvalues \cite{He:2010:J}.  

\item Estimate folds using $\prm{\mat{z}}\mapsto \prestL{\mu}{\tau}{m}$. 

\item Construct $\sqb{\mdft{r}}_\ell$ in \eqref{yft} for $\ell\in\left[ {0,N - 2} \right]$.

\item Estimate $\sqb{\mdft{\bs\gamma}}_\ell = \sqb{\mdft{y}}_\ell+ \sqb{\mdft{r}}_\ell$, $\ell \in \eset{N-1}$.

\item Using $\sqb{\mat{y}}_{\varphi,n} = \MO{\varphi\rob{nT}}$ implement Steps $1) \to 6)$ giving $\sqb{\mdft{\bs\gamma}}_\ell = \sqb{\mdft{\bs\varphi}}_\ell$. Construct matrix $\bs{\Psi}$ in \eqref{sest}.

\item Estimate $\sqb{\qsmat}_\ell, \ell \in \eset{N-1}$ via \eqref{sest}. %
Solve for $\{\rdc,h\}$ via $h\sqb{\ell}*\rob{\tilde{s} - \rdc \delta }\sqb{\ell} = 0 $ and define $\sqb{\hat{\mat{s}}}_0 = \rob{\sqb{\widetilde{\mat{s}}}_0 - \rdc}$.

\item Using $\sqb{\hat{\mat{s}}}$, estimate $K$ following Step 3). 

\item Estimate $s_K$ using $\prm{\hat{\mat{s}}} \mapsto \prest{c}{t}{k}$.

The $\rob{K+1}$ tap filter $\mat{h}$ is available via Step 8).
\end{enumerate}
\caption{Super-resolution from Modulo Samples ({\texttt{US}-\texttt{SR}})}
\label{alg:1}
\end{algorithm}

\bpara{Practical Aspects.} 
\begin{enumerate}[label = (\alph*), leftmargin = 30pt,itemsep = 5pt]
\item \emph{Noisy Scenario.} Real data may bear quantization and system noise, \eg Gaussian sources, see \fig{fig:demo}. We resort to oversampling \cite{Bhandari:2020:Ja} to tackle such adversarial effects. Prony's method is notoriously unstable with moderate noise \cite{Blu:2008:J,Batenkov:2013:J}, specially for high values of $\ML$. Therefore, we recommend using the matrix pencil method (\texttt{MPM}) \cite{Hua:1990:J} which is quite effective in practice; in \cite{Bhandari:2021:J} we have shown that with real data, as many as $\ML = 161$ folds can be estimated. 

\item \emph{Estimating the Number of Folds.} In our discussion we have assumed that $\ML$ is known. When $\bs{\gamma}$ is accessible, indeed $\ML$ is the sparsity of $\bar{\mat{r}} = \bs{\Delta}\rob{\bs{\gamma}-\mat{y}}$ \cite{Bhandari:2021:J}. In the absence of $\bs{\gamma}$, and in particular, when working with noisy data, estimating $\ML$ may turn out to be a difficult task. For the parametric form, ${\left[ {\mathbf{b}} \right]_\ell } = \sum\nolimits_{m = 0}^{M-1} {{\alpha_m}{e^{-\jmath {\nu_m}\ell }}}$, the case with $\mdft{r}$ and $\hat{\mat{s}}$, $M$ manifests as the rank of the Toeplitz or Hankel matrix constructed from $\mat{b}$ \cite{Figueiredo:1982:J,Blu:2008:J}. Rank thresholding based on the \emph{second order statistic} of the eigenvalues \cite{He:2010:J} is a reliable measure in that case. We use the same for estimating $K$ from data. Hence, our recovery method does not rely on the knowledge of $\lambda$, $K$ or $\ML$. Incorporating these practical aspects, we summarize the recovery approach in Alg.~\ref{alg:1}.
\end{enumerate}

\begin{figure}[!t]
\centering
     \scalebox{1.05}{\input{TikzFig}}  

\caption{Pipeline for hardware experiment, data is plotted in \fig{fig:exp}.}
\label{fig:HP}
\end{figure}
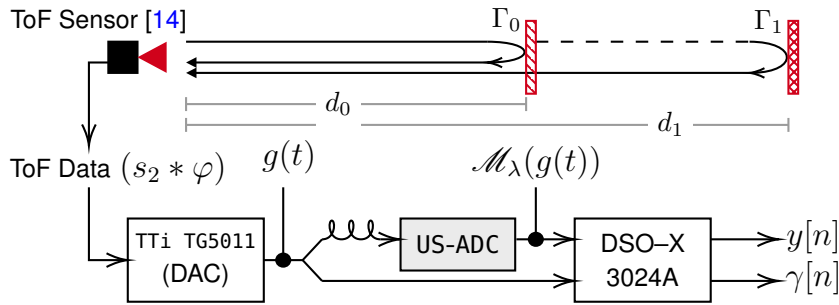

\begin{figure}[!t]
\centering
\begin{overpic}[width=0.65\columnwidth]{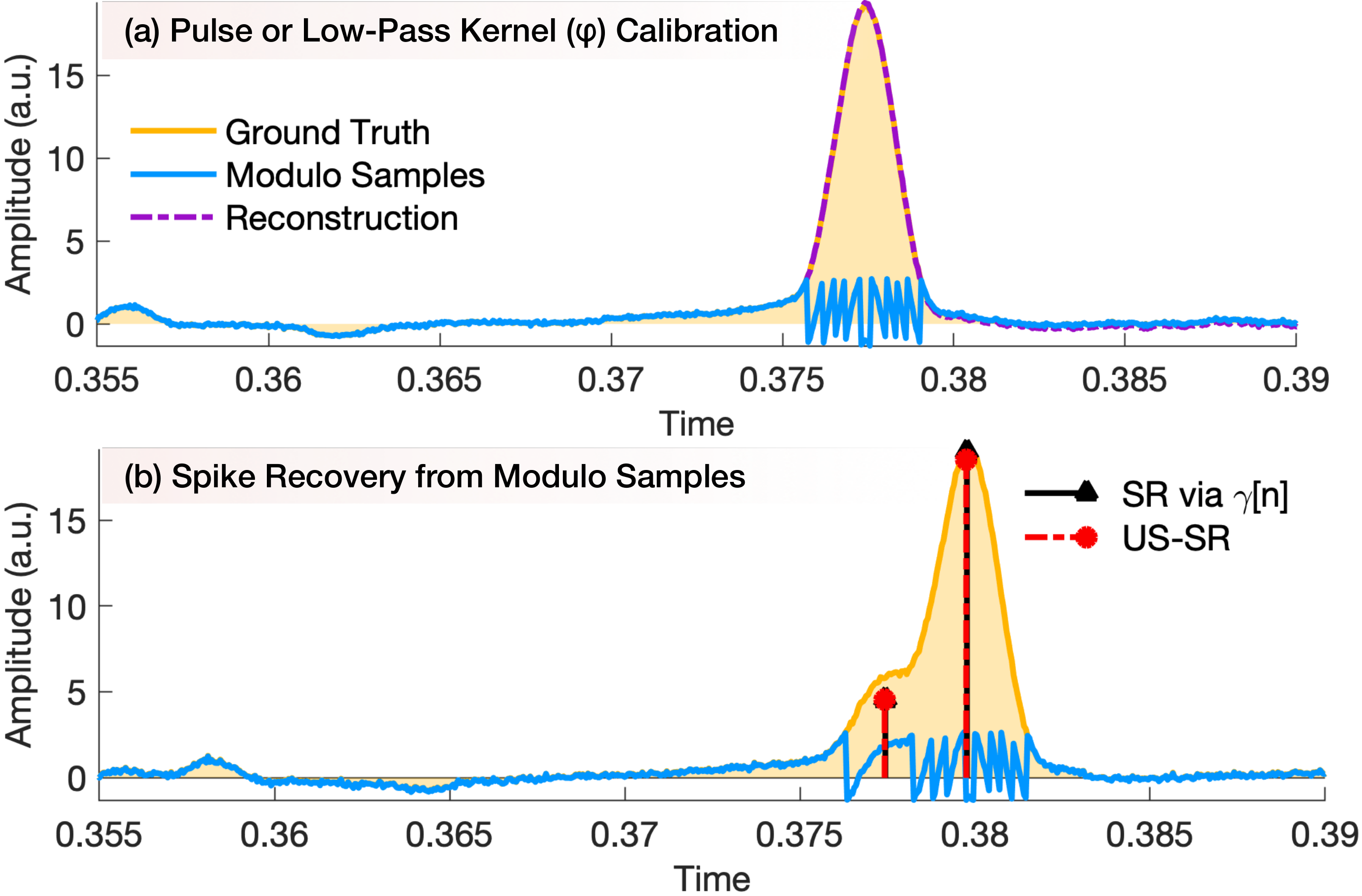}
\put (16.5,46) {{\sf{\bfseries\fontsize{7.5pt}{7.5pt}\selectfont {MSE} $=9.09\ep{-3}$}}}
\put (34,56) {{\sf{\bfseries\fontsize{7.5pt}{7.5pt}\selectfont $\pmb{\gamma\sqb{n}} = \pmb{\varphi\rob{nT}}$}}}
\put (38,52) {{\sf{\bfseries\fontsize{7.5pt}{7.5pt}\selectfont $\pmb{y\sqb{n}}$}}}
\put (8,23) {\sf \fontsize{7.5pt}{7.5pt}\selectfont 
$\begin{array}{*{20}{c}}
  {{c_0}}&{{c_1}}&{{t_0}} \mbox{ (ms)}&{{t_1}} \mbox{ (ms)} \\ 
  {4.54}&{18.96}&{377.431}&{379.769} 
\end{array}$}
\put (8,16) {\sf\fontsize{7.5pt}{7.5pt}\selectfont 
$\begin{array}{*{20}{c}}
  {{{\widetilde c}_0}}&{{{\widetilde c}_1}}&{{{\widetilde t}_0}}  \mbox{ (ms)} &{{{\widetilde t}_1}}   \mbox{ (ms)}\\ 
  {4.54}&{18.52}&{377.411}&{379.748} 
\end{array}$
}
\end{overpic}
\caption{Hardware Experiment for Sparse Super-resolution. (a) Conventional and modulo samples of the kernel $\varphi\rob{t}$. (b) Single-pixel data $\rob{\gamma\sqb{n}}$ from ToF sensor \cite{Bhandari:2016:J}, corresponding modulo samples $\rob{y\sqb{n}}$ and recovered spikes $\{c_k,t_k\}$ via ${\gamma\sqb{n}}$ (ground truth) and $\{\widetilde{c}_k,\widetilde{t}_k\}$ via ${y\sqb{n}}$ (Alg.~\ref{alg:1}), respectively.  }
\label{fig:exp}
\end{figure}
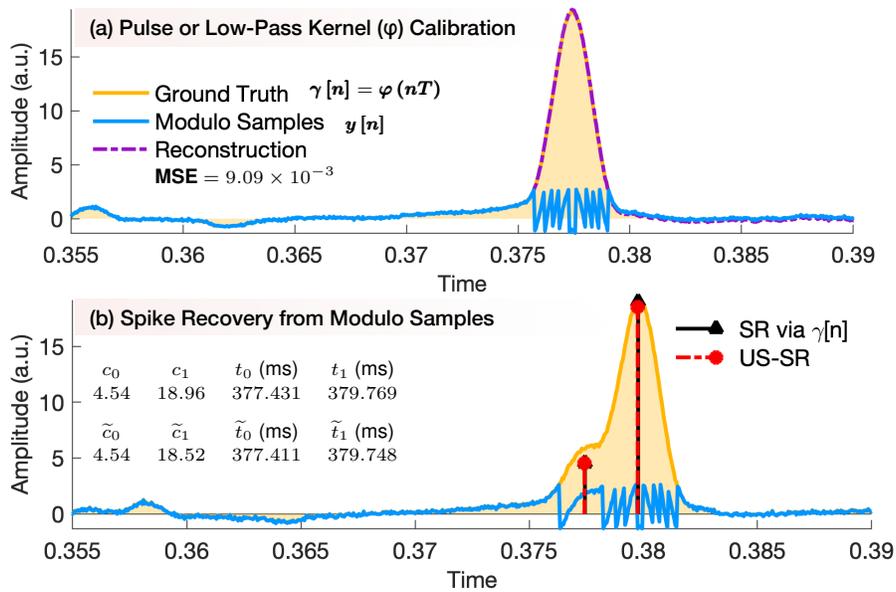

\section{Hardware Based Experimental Validation}
\label{sec:HEV}
The \ussr~method in Alg.~\ref{alg:1} performs up to machine precision with computer simulations. To demonstrate the practical effectivity of our method, specially in the presence of quantization and system noise, we setup the acquisition pipeline shown in \fig{fig:HP}. Samples of $K = 2$ sparse signal are obtained via a time-of-flight (ToF) imaging sensor (also cf.~Fig.~7, \cite{Bhandari:2016:J}). In particular, objects with albedo $c_k \propto \Gamma_k$ at depths $t_k = 2d_k/\nu$ (cf.~\fig{fig:HP}) where $\nu = 3\times10^8$ m/s (speed of light), result in $s_2\rob{t}$ \eqref{sod}. The ToF imager measures \cite{Bhandari:2016:Cb,Bhandari:2020:J} low-pass projections of $s_2\rob{t}$ via $\varphi\rob{t}$ which can be calibrated. We convert ToF and pulse $\rob{\varphi}$ samples into a continuous-time (CT) signal using \texttt{TTi TG5011} waveform generator (cf.~\fig{fig:HP}). The CT output is split into $2$ channels fed to the \texttt{DSO}-\texttt{X} \texttt{3024A} oscilloscope with inbuilt ADC, thus yielding \textit{(a)} $y\sqb{n}$, \usadc~based modulo samples and \textit{(b)} $\gamma\sqb{n}$, the conventional samples (our ground truth). With $\lambda = \nicefrac{202}{100}$ and $T = 50 \ \mu\mathrm{s}$ we obtain $N= 1000$ samples of $y\sqb{n}$ and $\gamma\sqb{n}$, corresponding to $\varphi\rob{t}$ (cf.~\fig{fig:exp}(a)) and $g\rob{t}$ (cf.~\fig{fig:exp}(b)), respectively. This gives $\tau = 50\ \mathrm{ms}$ and we estimate $P = 23$ from $\{\varphi\rob{nT}\}_n$. However, for Step 8) in Alg.~\ref{alg:1}, we use $\sqb{\qmat}_p = \widehat{s}_p, |p| \leq P' = 20$ as the samples on the extreme $|p|>20$ are deemed noisy. {\bf Kernel Calibration and Reconstruction.} Note that $\max |\varphi\rob{nT}|\approx 9.52\lambda$. Using Step 3) of Alg.~\ref{alg:1}, we estimate $\ML_{,\varphi} = 10$ which is consistent with the sparsity of  $\bar{\mat{r}}=\bs{\Delta}\rob{\bs{\gamma}-\mat{y}}$. Since the data in \fig{fig:exp}(a) can be interpreted as filtering with $K = 1$ spike, we can use Step 7) in Alg.~\ref{alg:1} to estimate $\varphi$ 
from $y_\varphi\sqb{n} = \MO{\varphi\rob{nT}}$ (also see \cite{Bhandari:2021:J}). Although not required, the reconstructed signal $\widetilde\varphi$ is shown in \fig{fig:exp}(a) and the resulting mean-squared error (MSE), \ie~$\EuScript{E}\rob{\widetilde\varphi,\varphi} = \frac{1}{N}\sum\nolimits_{n=0}^{N-1} \vert \varphi\rob{nT} - \widetilde\varphi\rob{nT} \vert^2$ is $9.09\ep{-3}$. 
{\bf Spike Recovery.} The data is plotted in \fig{fig:exp}(b). Given $\gamma\sqb{n}$, we observe $\max |g\rob{nT}|\approx 9.46\lambda$. Using Step 3) of Alg.~\ref{alg:1}, we estimate $\ML_{,g} = 10$. Using $\{\gamma\sqb{n}\}_{n=0}^{N-1}$ we estimate $\{c_k,t_k\}$ using conventional super-resolution methods \cite{Bhandari:2016:J,Bhandari:2016:Cb,Bhandari:2014:Cb} rounding $\{c_k\}_k$ and $\{t_k\}_k$ to $2$ and $6$ decimals, respectively. Because these values are consistent with the ToF imaging experimental parameters \cite{Bhandari:2016:J,Bhandari:2014:Cb}, this serves as our ground truth for $s_2\rob{t}$. Thereon, we use Alg.~\ref{alg:1} to estimate the spikes via modulo samples. The results are tabulated below.
\begin{table}[!h]
\centering
\resizebox{0.45\textwidth}{!}{%
\begin{tabular}{@{}lcccc @{}}
\toprule
{}	& 	${c_0}$	&	${c_1}$	&	{${t_0}$} {(ms)}		&	{${t_1}$} { (ms)}  \\ \midrule
Ground Truth via ${\gamma \left[ n \right]}$	& 	$4.54$ 	& 	$18.96$ 	&	 $377.431$ 		& 	$379.769$ \\ 
Using $y\sqb{n}$ and Alg.~\ref{alg:1} & $4.54$ & $18.52$ & $377.411$ & $379.748$
\\ \bottomrule
\end{tabular}%
}
\end{table}

\noindent The output of \ussr~is shown with the ground truth in \fig{fig:exp}(b). The worst case absolute error, ${\max _k}\rob{|{c_k} - {\widetilde c_k}|}\propto \ep{-2}$ and ${\max _k}\rob{|{t_k} - {\widetilde t_k}|}\propto \ep{-4}$ (sec), respectively, demonstrates the effectivity of \ussr~in a realistic setting.

\section{Conclusions}
We considered the problem of recovering sparse signals from low-pass filtered, modulo samples. By leveraging that modulo folds and the sparse signal result in an intertwined, doubly sparse structure and observing that they map to trigonometric polynomials in the Fourier domain, we developed an exact super-resolution approach, namely \ussr. Our recovery guarantee purely depends on the input signal sparsity and the number of folds, thus offering a practically amenable bound compared to \cite{Bhandari:2018:Ca}. The \ussr~approach is agnostic to the ADC threshold $\rob{\lambda}$ and relies only on first order difference; this latter aspect avoids instabilities arising from higher order differences \cite{Bhandari:2018:Ca}, thus offering a practically attractive solution. Hardware experiments with our modulo ADC validate the performance of \ussr. In the context of the USF, development of robust SR algorithms and performance analysis in the presence of noise remain interesting topics for future exploration.

\bpara{Acknowledgment.} The author thanks the reviewers for their encouraging remarks, in particular, one of the reviewers who alluded to ``Back in the U.S.S.R.'' from \emph{The Beatles} which was unknown to the author and inspired the title of this paper.

\ifCLASSOPTIONcaptionsoff
\newpage
\fi

\newpage

\end{document}

%% file: TikzFig.tex
 
\tikzset{
pattern size/.store in=\mcSize, 
pattern size = 5pt,
pattern thickness/.store in=\mcThickness, 
pattern thickness = 0.3pt,
pattern radius/.store in=\mcRadius, 
pattern radius = 1pt}
\makeatletter
\pgfutil@ifundefined{pgf@pattern@name@_ai82tgxfz}{
\pgfdeclarepatternformonly[\mcThickness,\mcSize]{_ai82tgxfz}
{\pgfqpoint{0pt}{0pt}}
{\pgfpoint{\mcSize}{\mcSize}}
{\pgfpoint{\mcSize}{\mcSize}}
{
\pgfsetcolor{\tikz@pattern@color}
\pgfsetlinewidth{\mcThickness}
\pgfpathmoveto{\pgfqpoint{0pt}{\mcSize}}
\pgfpathlineto{\pgfpoint{\mcSize+\mcThickness}{-\mcThickness}}
\pgfpathmoveto{\pgfqpoint{0pt}{0pt}}
\pgfpathlineto{\pgfpoint{\mcSize+\mcThickness}{\mcSize+\mcThickness}}
\pgfusepath{stroke}
}}
\makeatother

 
\tikzset{
pattern size/.store in=\mcSize, 
pattern size = 5pt,
pattern thickness/.store in=\mcThickness, 
pattern thickness = 0.3pt,
pattern radius/.store in=\mcRadius, 
pattern radius = 1pt}
\makeatletter
\pgfutil@ifundefined{pgf@pattern@name@_ypggecerl}{
\pgfdeclarepatternformonly[\mcThickness,\mcSize]{_ypggecerl}
{\pgfqpoint{0pt}{-\mcThickness}}
{\pgfpoint{\mcSize}{\mcSize}}
{\pgfpoint{\mcSize}{\mcSize}}
{
\pgfsetcolor{\tikz@pattern@color}
\pgfsetlinewidth{\mcThickness}
\pgfpathmoveto{\pgfqpoint{0pt}{\mcSize}}
\pgfpathlineto{\pgfpoint{\mcSize+\mcThickness}{-\mcThickness}}
\pgfusepath{stroke}
}}
\makeatother
\tikzset{every picture/.style={line width=0.75pt}} 

\begin{tikzpicture}[x=0.70pt,y=0.75pt,yscale=-1,xscale=1]
\path (180,55);

\draw   (50,130) -- (120,130) -- (120,170) -- (50,170) -- cycle ;
\draw   (180,140) .. controls (173,140) and (173,131) .. (175,131) .. controls (177,131) and (177,140) .. (170,140) .. controls (163,140) and (163,131) .. (165,131) .. controls (167,131) and (167,140) .. (160,140) .. controls (153,140) and (153,131) .. (155,131) .. controls (157,131) and (157,140) .. (150,140) ;
\draw    (48,150) -- (30,150) -- (30,115) ;
\draw [shift={(50,150)}, rotate = 180] [color={rgb, 255:red, 0; green, 0; blue, 0 }  ][line width=0.75]    (10.93,-3.29) .. controls (6.95,-1.4) and (3.31,-0.3) .. (0,0) .. controls (3.31,0.3) and (6.95,1.4) .. (10.93,3.29)   ;
\draw   (280,130) -- (350,130) -- (350,170) -- (280,170) -- cycle ;
\draw    (150,160) -- (278,160) ;
\draw [shift={(280,160)}, rotate = 180] [color={rgb, 255:red, 0; green, 0; blue, 0 }  ][line width=0.75]    (10.93,-3.29) .. controls (6.95,-1.4) and (3.31,-0.3) .. (0,0) .. controls (3.31,0.3) and (6.95,1.4) .. (10.93,3.29)   ;
\draw    (250,140) -- (278,140) ;
\draw [shift={(280,140)}, rotate = 180] [color={rgb, 255:red, 0; green, 0; blue, 0 }  ][line width=0.75]    (10.93,-3.29) .. controls (6.95,-1.4) and (3.31,-0.3) .. (0,0) .. controls (3.31,0.3) and (6.95,1.4) .. (10.93,3.29)   ;
\draw    (120,150) -- (140,150) -- (150,140) ;
\draw    (140,150) -- (150,160) ;
\draw    (180,140) -- (188,140) ;
\draw [shift={(190,140)}, rotate = 180] [color={rgb, 255:red, 0; green, 0; blue, 0 }  ][line width=0.75]    (10.93,-3.29) .. controls (6.95,-1.4) and (3.31,-0.3) .. (0,0) .. controls (3.31,0.3) and (6.95,1.4) .. (10.93,3.29)   ;
\draw    (130,150) -- (130,115) ;
\draw [shift={(130,150)}, rotate = 270] [color={rgb, 255:red, 0; green, 0; blue, 0 }  ][fill={rgb, 255:red, 0; green, 0; blue, 0 }  ][line width=0.75]      (0, 0) circle [x radius= 3.35, y radius= 3.35]   ;
\draw    (260,140) -- (260,115) ;
\draw [shift={(260,140)}, rotate = 270] [color={rgb, 255:red, 0; green, 0; blue, 0 }  ][fill={rgb, 255:red, 0; green, 0; blue, 0 }  ][line width=0.75]      (0, 0) circle [x radius= 3.35, y radius= 3.35]   ;
\draw    (350,140) -- (383,140) ;
\draw [shift={(385,140)}, rotate = 180] [color={rgb, 255:red, 0; green, 0; blue, 0 }  ][line width=0.75]    (10.93,-3.29) .. controls (6.95,-1.4) and (3.31,-0.3) .. (0,0) .. controls (3.31,0.3) and (6.95,1.4) .. (10.93,3.29)   ;
\draw    (350,160) -- (383,160) ;
\draw [shift={(385,160)}, rotate = 180] [color={rgb, 255:red, 0; green, 0; blue, 0 }  ][line width=0.75]    (10.93,-3.29) .. controls (6.95,-1.4) and (3.31,-0.3) .. (0,0) .. controls (3.31,0.3) and (6.95,1.4) .. (10.93,3.29)   ;
\draw  [fill={rgb, 255:red, 0; green, 0; blue, 0 }  ,fill opacity=1 ] (40,45) -- (55,45) -- (55,60) -- (40,60) -- cycle ;
\draw  [draw opacity=0][fill={rgb, 255:red, 208; green, 2; blue, 27 }  ,fill opacity=1 ] (55,52.5) -- (70,45) -- (70,60) -- cycle ;

\draw [line width=0.75]    (235,45) .. controls (260.11,45.06) and (260.43,54.56) .. (236.86,54.99) ;
\draw [shift={(235,55)}, rotate = 360.13] [color={rgb, 255:red, 0; green, 0; blue, 0 }  ][line width=0.75]    (7.65,-2.3) .. controls (4.86,-0.97) and (2.31,-0.21) .. (0,0) .. controls (2.31,0.21) and (4.86,0.98) .. (7.65,2.3)   ;
\draw [line width=0.75]    (370,45) .. controls (395.11,45.09) and (395.43,59.35) .. (371.86,59.98) ;
\draw [shift={(370,60)}, rotate = 360.2] [color={rgb, 255:red, 0; green, 0; blue, 0 }  ][line width=0.75]    (7.65,-2.3) .. controls (4.86,-0.97) and (2.31,-0.21) .. (0,0) .. controls (2.31,0.21) and (4.86,0.98) .. (7.65,2.3)   ;
\draw    (80,45) -- (235,45) ;
\draw    (83,60) -- (370,60) ;
\draw [shift={(80,60)}, rotate = 360] [fill={rgb, 255:red, 0; green, 0; blue, 0 }  ][line width=0.08]  [draw opacity=0] (5.36,-2.57) -- (0,0) -- (5.36,2.57) -- cycle    ;
\draw  [color={rgb, 255:red, 208; green, 2; blue, 27 }  ,draw opacity=1 ][pattern=_ai82tgxfz,pattern size=3.75pt,pattern thickness=0.75pt,pattern radius=0pt, pattern color={rgb, 255:red, 208; green, 2; blue, 27}] (390,35) -- (395,35) -- (395,70) -- (390,70) -- cycle ;
\draw    (83,55) -- (235,55) ;
\draw [shift={(80,55)}, rotate = 360] [fill={rgb, 255:red, 0; green, 0; blue, 0 }  ][line width=0.08]  [draw opacity=0] (5.36,-2.57) -- (0,0) -- (5.36,2.57) -- cycle    ;
\draw    (30,93) -- (30,55) -- (40,55) ;
\draw [shift={(30,95)}, rotate = 270] [color={rgb, 255:red, 0; green, 0; blue, 0 }  ][line width=0.75]    (10.93,-3.29) .. controls (6.95,-1.4) and (3.31,-0.3) .. (0,0) .. controls (3.31,0.3) and (6.95,1.4) .. (10.93,3.29)   ;
\draw  [dash pattern={on 4.5pt off 4.5pt}]  (260,45) -- (370,45) ;
\draw [color={rgb, 255:red, 155; green, 155; blue, 155 }  ,draw opacity=1 ]   (80,75) -- (255,75) ;
\draw [shift={(255,75)}, rotate = 180] [color={rgb, 255:red, 155; green, 155; blue, 155 }  ,draw opacity=1 ][line width=0.75]    (0,3.35) -- (0,-3.35)   ;
\draw [shift={(80,75)}, rotate = 180] [color={rgb, 255:red, 155; green, 155; blue, 155 }  ,draw opacity=1 ][line width=0.75]    (0,3.35) -- (0,-3.35)   ;
\draw [color={rgb, 255:red, 155; green, 155; blue, 155 }  ,draw opacity=1 ]   (80,85) -- (390,85) ;
\draw [shift={(390,85)}, rotate = 180] [color={rgb, 255:red, 155; green, 155; blue, 155 }  ,draw opacity=1 ][line width=0.75]    (0,3.35) -- (0,-3.35)   ;
\draw [shift={(80,85)}, rotate = 180] [color={rgb, 255:red, 155; green, 155; blue, 155 }  ,draw opacity=1 ][line width=0.75]    (0,3.35) -- (0,-3.35)   ;
\draw  [color={rgb, 255:red, 208; green, 2; blue, 27 }  ,draw opacity=1 ][pattern=_ypggecerl,pattern size=3.75pt,pattern thickness=0.75pt,pattern radius=0pt, pattern color={rgb, 255:red, 208; green, 2; blue, 27}] (255,35) -- (260,35) -- (260,70) -- (255,70) -- cycle ;

\draw (85,150) node  [font=\small] [align=left] {\begin{minipage}[lt]{55pt}\setlength\topsep{0pt}
\begin{center}
{ \texttt{TTi TG5011}}\\ {\sf (DAC)}
\end{center}

\end{minipage}};
\draw  [fill={rgb, 255:red, 155; green, 155; blue, 155 }  ,fill opacity=0.2 ]  (190.38,130) -- (249.38,130) -- (249.38,155) -- (190.38,155) -- cycle  ;
\draw (219.88,142.5) node   [align=left] {{\usadc}};
\draw (315,150) node  [font=\small] [align=left] {\begin{minipage}[lt]{31.79pt}\setlength\topsep{0pt}
\begin{center}
{\sf DSO--X}\\{\sf 3024A}
\end{center}

\end{minipage}};
\draw (133,100.5) node  [font=\large]  {$g( t)$};
\draw (403,138.5) node  [font=\large]  {$y[ n]$};
\draw (403,158.5) node  [font=\large]  {$\gamma [ n]$};
\draw (45,106) node  [font=\small] [align=left] {{\sf ToF Data} \large$\rob{s_2*\varphi}$};
\draw (245,34.5) node    {$\Gamma _{0}$};
\draw (380,35.5) node    {$\Gamma _{1}$};
\draw  [draw opacity=0][fill={rgb, 255:red, 255; green, 255; blue, 255 }  ,fill opacity=1 ]  (147,63) -- (170,63) -- (170,89) -- (147,89) -- cycle  ;
\draw (150,67.4) node [anchor=north west][inner sep=0.75pt]  [font=\normalsize]  {$d_{0}$};
\draw  [draw opacity=0][fill={rgb, 255:red, 255; green, 255; blue, 255 }  ,fill opacity=1 ]  (318,73) -- (341,73) -- (341,99) -- (318,99) -- cycle  ;
\draw (321,77.4) node [anchor=north west][inner sep=0.75pt]  [font=\normalsize]  {$d_{1}$};
\draw (35,35) node  [font=\small] [align=left] {\sf ToF Sensor \cite{Bhandari:2016:J}};
\draw  [draw opacity=0]  (225,90) -- (296,90) -- (296,116) -- (225,116) -- cycle  ;
\draw (260.5,103) node  [font=\large] [align=left] {$\displaystyle \mathscr{M}_{\lambda }( g( t))$};
\end{tikzpicture}

%% file: SPL-32939-2022_Report.bbl
\begin{thebibliography}{10}
\providecommand{\url}[1]{#1}
\csname url@samestyle\endcsname
\providecommand{\newblock}{\relax}
\providecommand{\bibinfo}[2]{#2}
\providecommand{\BIBentrySTDinterwordspacing}{\spaceskip=0pt\relax}
\providecommand{\BIBentryALTinterwordstretchfactor}{4}
\providecommand{\BIBentryALTinterwordspacing}{\spaceskip=\fontdimen2\font plus
\BIBentryALTinterwordstretchfactor\fontdimen3\font minus
  \fontdimen4\font\relax}
\providecommand{\BIBforeignlanguage}[2]{{%
\expandafter\ifx\csname l@#1\endcsname\relax
\typeout{** WARNING: IEEEtran.bst: No hyphenation pattern has been}%
\typeout{** loaded for the language `#1'. Using the pattern for}%
\typeout{** the default language instead.}%
\else
\language=\csname l@#1\endcsname
\fi
#2}}
\providecommand{\BIBdecl}{\relax}
\BIBdecl

\bibitem{Bhandari:2017:C}
\BIBentryALTinterwordspacing
A.~Bhandari, F.~Krahmer, and R.~Raskar, ``On unlimited sampling,'' in
  \emph{Intl. Conf. on Sampling Theory and Applications (SampTA)}, Jul. 2017.
\BIBentrySTDinterwordspacing

\bibitem{Bhandari:2018:Ca}
------, ``Unlimited sampling of sparse signals,'' in \emph{{IEEE} Intl. Conf.
  on Acoustics, Speech and Signal Processing (ICASSP)}, Apr. 2018.

\bibitem{Bhandari:2018:C}
------, ``Unlimited sampling of sparse sinusoidal mixtures,'' in \emph{{IEEE}
  Intl. Sym. on Information Theory ({ISIT})}, Jun. 2018.

\bibitem{Bhandari:2019:C}
A.~Bhandari and F.~Krahmer, ``On identifiability in unlimited sampling,'' in
  \emph{Intl. Conf. on Sampling Theory and Applications (SampTA)}, Jul. 2019.

\bibitem{Bhandari:2020:Pata}
A.~Bhandari, F.~Krahmer, and R.~Raskar, ``Methods and apparatus for modulo
  sampling and recovery,'' US Patent US10\,651\,865B2, May, 2020.

\bibitem{Bhandari:2020:Ja}
------, ``On unlimited sampling and reconstruction,'' \emph{{IEEE} Trans. Sig.
  Proc.}, vol.~69, pp. 3827--3839, Dec. 2020.

\bibitem{Bhandari:2021:J}
A.~Bhandari, F.~Krahmer, and T.~Poskitt, ``Unlimited sampling from theory to
  practice: {Fourier}-{Prony} recovery and prototype {ADC},'' \emph{{IEEE}
  Trans. Sig. Proc.}, Sep. 2021.

\bibitem{FernandezMenduina:2021:J}
S.~Fernandez-Menduina, F.~Krahmer, G.~Leus, and A.~Bhandari, ``Computational
  array signal processing via modulo non-linearities,'' \emph{{IEEE} Trans.
  Sig. Proc.}, Jul. 2021.

\bibitem{Florescu:2022:J}
D.~Florescu, F.~Krahmer, and A.~Bhandari, ``The surprising benefits of
  hysteresis in unlimited sampling: {Theory}, algorithms and experiments,''
  \emph{{IEEE} Trans. Sig. Proc.}, vol.~70, pp. 616--630, 2022.

\bibitem{Abel:1991:C}
\BIBentryALTinterwordspacing
J.~Abel and J.~Smith, ``Restoring a clipped signal,'' in \emph{{IEEE} Intl.
  Conf. on Acoustics, Speech and Sig. Proc. ({ICASSP})}, 1991.
\BIBentrySTDinterwordspacing

\bibitem{Esqueda:2016:J}
\BIBentryALTinterwordspacing
F.~Esqueda, S.~Bilbao, and V.~Valimaki, ``Aliasing reduction in clipped
  signals,'' \emph{{IEEE} Trans. Sig. Proc.}, vol.~64, no.~20, pp. 5255--5267,
  Oct. 2016.
\BIBentrySTDinterwordspacing

\bibitem{Olofsson:2005}
\BIBentryALTinterwordspacing
T.~Olofsson, ``Deconvolution and model-based restoration of clipped ultrasonic
  signals,'' \emph{{IEEE} Trans. Instrum. Meas.}, vol.~54, no.~3, pp.
  1235--1240, Jun. 2005.
\BIBentrySTDinterwordspacing

\bibitem{Bhandari:2022:Book}
A.~Bhandari, A.~Kadambi, and R.~Raskar, \emph{Computational Imaging},
  1st~ed.\hskip 1em plus 0.5em minus 0.4em\relax MIT Press, Jun. 2022, \newline
  Open Access URL: https://imagingtext.github.io/.

\bibitem{Bhandari:2016:J}
A.~Bhandari and R.~Raskar, ``Signal processing for {Time-of-Flight} imaging
  sensors: {A}n introduction to inverse problems in computational {3-D}
  imaging,'' \emph{{IEEE} Signal Process. Mag.}, vol.~33, no.~5, pp. 45--58,
  Sep. 2016.

\bibitem{Bhandari:2020:C}
A.~Bhandari and F.~Krahmer, ``{HDR} imaging from quantization noise,'' in
  \emph{{IEEE} Intl. Conf. on Image Processing ({ICIP})}, Oct. 2020, pp.
  101--105.

\bibitem{Romanov:2019:J}
E.~Romanov and O.~Ordentlich, ``Above the {Nyquist} rate, modulo folding does
  not hurt,'' \emph{{IEEE} Signal Process. Lett.}, vol.~26, no.~8, pp.
  1167--1171, Aug. 2019.

\bibitem{Musa:2018:C}
O.~Musa, P.~Jung, and N.~Goertz, ``Generalized approximate message passing for
  unlimited sampling of sparse signals,'' in \emph{{IEEE} Global Conf. on Sig.
  and Info. Proc. ({GlobalSIP})}.\hskip 1em plus 0.5em minus 0.4em\relax
  {IEEE}, Nov. 2018.

\bibitem{Rudresh:2018:C}
\BIBentryALTinterwordspacing
S.~Rudresh, A.~Adiga, B.~A. Shenoy, and C.~S. Seelamantula, ``Wavelet-based
  reconstruction for unlimited sampling,'' in \emph{{IEEE} Intl. Conf. on
  Acoustics, Speech and Sig. Proc. ({ICASSP})}, Apr. 2018.
\BIBentrySTDinterwordspacing

\bibitem{Gan:2020:C}
L.~Gan and H.~Liu, ``High dynamic range sensing using multi-channel modulo
  samplers,'' in \emph{{IEEE} Sensor Array and Multichannel Sig. Proc. Workshop
  {(SAM)}}, Jun. 2020.

\bibitem{Shah:2021:J}
V.~Shah and C.~Hegde, ``Sparse signal recovery from modulo observations,''
  \emph{{EURASIP} Journal on Advances in Signal Processing}, vol. 2021, no.~1,
  Apr. 2021.

\bibitem{Prasanna:2021:J}
D.~Prasanna, C.~Sriram, and C.~R. Murthy, ``On the identifiability of sparse
  vectors from modulo compressed sensing measurements,'' \emph{{IEEE} Signal
  Process. Lett.}, vol.~28, pp. 131--134, Jan. 2021.

\bibitem{Figueiredo:1982:J}
R.~J. P.~D. Figueiredo and C.-L. Hu, ``Waveform feature extraction based on
  {Tauberian} approximation,'' \emph{{IEEE} Trans. Pattern Anal. Mach.
  Intell.}, vol. {PAMI}-4, no.~2, pp. 105--116, Mar. 1982.

\bibitem{Kirsteins:1987}
I.~Kirsteins, ``High resolution time delay estimation,'' in \emph{{IEEE} Intl.
  Conf. on Acoustics, Speech and Signal Processing (ICASSP)}, Apr. 1987.

\bibitem{Fuchs:1999:J}
J.-J. Fuchs, ``Multipath time-delay detection and estimation,'' \emph{{IEEE}
  Trans. Sig. Proc.}, vol.~47, no.~1, pp. 237--243, Jan. 1999.

\bibitem{Gedalyahu:2010:J}
K.~Gedalyahu and Y.~C. Eldar, ``Time-delay estimation from low-rate samples: A
  union of subspaces approach,'' \emph{{IEEE} Trans. Sig. Proc.}, vol.~58,
  no.~6, pp. 3017--3031, Jun. 2010.

\bibitem{Donoho:1992:J}
D.~L. Donoho, ``Superresolution via sparsity constraints,'' \emph{{SIAM}
  Journal on Mathematical Analysis}, vol.~23, no.~5, pp. 1309--1331, Sep. 1992.

\bibitem{Candes:2013:J}
E.~J. Cand{\`{e}}s and C.~Fernandez-Granda, ``Towards a mathematical theory of
  super-resolution,'' \emph{Communications on Pure and Applied Mathematics},
  vol.~67, no.~6, pp. 906--956, Apr. 2013.

\bibitem{Bhaskar:2013:J}
B.~N. Bhaskar, G.~Tang, and B.~Recht, ``Atomic norm denoising with applications
  to line spectral estimation,'' \emph{{IEEE} Trans. Sig. Proc.}, vol.~61,
  no.~23, pp. 5987--5999, Dec. 2013.

\bibitem{Li:2000:J}
L.~Li and T.~P. Speed, ``Parametric deconvolution of positive spike trains,''
  \emph{The Annals of Statistics}, vol.~28, no.~5, Oct. 2000.

\bibitem{Vetterli:2002:J}
M.~Vetterli, P.~Marziliano, and T.~Blu, ``Sampling signals with finite rate of
  innovation,'' \emph{{IEEE} Trans. Sig. Proc.}, vol.~50, no.~6, pp.
  1417--1428, Jun. 2002.

\bibitem{Blu:2008:J}
T.~Blu, P.-L. Dragotti, M.~Vetterli, P.~Marziliano, and L.~Coulot, ``Sparse
  sampling of signal innovations,'' \emph{{IEEE} Signal Process. Mag.},
  vol.~25, no.~2, pp. 31--40, Mar. 2008.

\bibitem{Bhandari:2014:Cb}
\BIBentryALTinterwordspacing
A.~Bhandari, A.~Kadambi, and R.~Raskar, ``Sparse linear operator identification
  without sparse regularization? {A}pplications to mixed pixel problem in
  time-of-flight range imaging,'' in \emph{{IEEE} Intl. Conf. on Acoustics,
  Speech and Signal Processing (ICASSP)}, May 2014.
\BIBentrySTDinterwordspacing

\bibitem{Bhandari:2016:Cb}
\BIBentryALTinterwordspacing
A.~Bhandari, A.~M. Wallace, and R.~Raskar, ``Super-resolved time-of-flight
  sensing via {FRI} sampling theory,'' in \emph{{IEEE} Intl. Conf. on
  Acoustics, Speech and Signal Processing (ICASSP)}, Mar. 2016.
\BIBentrySTDinterwordspacing

\bibitem{Bhandari:2017:Ca}
A.~Bhandari and T.~Blu, ``{FRI} sampling and time-varying pulses: {S}ome theory
  and four short stories,'' in \emph{{IEEE} Intl. Conf. on Acoustics, Speech
  and Signal Processing (ICASSP)}, Mar. 2017, pp. 3804--3808.

\bibitem{Bhandari:2020:J}
A.~Bhandari, M.~H. Conde, and O.~Loffeld, ``One-bit time-resolved imaging,''
  \emph{{IEEE} Trans. Pattern Anal. Mach. Intell.}, vol.~42, no.~7, pp.
  1630--1641, Jul. 2020.

\bibitem{Wolf:1983:J}
J.~Wolf, ``Redundancy, the discrete {F}ourier transform, and impulse noise
  cancellation,'' \emph{{IEEE} Trans. Commun.}, vol.~31, no.~3, pp. 458--461,
  Mar. 1983.

\bibitem{Batenkov:2013:J}
D.~Batenkov and Y.~Yomdin, ``On the accuracy of solving confluent {Prony}
  systems,'' \emph{SIAM Journal on Applied Mathematics}, vol.~73, no.~1, pp.
  134--154, Jan. 2013.

\bibitem{Hua:1990:J}
Y.~Hua and T.~K. Sarkar, ``Matrix pencil method for estimating parameters of
  exponentially damped/undamped sinusoids in noise,'' \emph{{IEEE} Trans.
  Acoust., Speech, Signal Process.}, vol.~38, no.~5, pp. 814--824, May 1990.

\bibitem{He:2010:J}
Z.~He, A.~Cichocki, S.~Xie, and K.~Choi, ``Detecting the number of clusters in
  n-way probabilistic clustering,'' \emph{{IEEE} Trans. Pattern Anal. Mach.
  Intell.}, vol.~32, no.~11, pp. 2006--2021, Nov. 2010.

\end{thebibliography}
